\documentclass[12pt,preprint]{aastex}

\def\etal{et~al.}
\def\hst{{\it HST}}

\shorttitle{ACS VCS IV: Data Reduction Procedures for SBF}
\shortauthors{Mei et al.}

\begin{document}


\title{The ACS Virgo Cluster Survey IV: Data Reduction Procedures for Surface
Brightness Fluctuation Measurements with the Advanced Camera for Surveys}


\author{Simona Mei\altaffilmark{1}, John P. Blakeslee\altaffilmark{1}, John L. Tonry\altaffilmark{2},  Andr\'es Jord\'an\altaffilmark{3,4,5},  Eric W. Peng\altaffilmark{3}, Patrick C\^ot\'e\altaffilmark{3}, Laura Ferrarese\altaffilmark{3},  David~Merritt\altaffilmark{6}, Milo\v s Milosavljevi\'c \altaffilmark{7,8},\\ and Michael J. West\altaffilmark{9}}


\altaffiltext{1}{Department of Physics and Astronomy, Johns Hopkins University, Baltimore, MD 21218; smei@pha.jhu.edu, jpb@pha.jhu.edu}
\altaffiltext{2}{Institute of Astronomy, University of Hawaii, 2680 Woodlawn Drive, Honolulu, HI 96822; jt@ifa.hawaii.edu}
\altaffiltext{3}{Department of Physics and Astronomy, Rutgers University, Piscataway, NJ 08854; andresj@physics.rutgers.edu, ericpeng@physics.rutgers.edu, pcote@physics.rutgers.edu, lff@physics.rutgers.edu}
\altaffiltext{4}{Astrophysics, Denys Wilkinson Building, University of Oxford, 1 Keble Road, Oxford, OX1 3RH, UK}
\altaffiltext{5}{Claudio Anguita Fellow}
\altaffiltext{6}{Department of Physics, Rochester Institute of Technology, 84 Lomb Memorial Drive, Rochester, NY 14623; merritt@cis.rit.edu}
\altaffiltext{7}{Theoretical Astrophysics, California Institute of Technology, Pasadena, CA 91125; milos@tapir.caltech.edu}
\altaffiltext{8}{Sherman M. Fairchild Fellow}
\altaffiltext{9}{Department of Physics and Astronomy, University of Hawaii, Hilo, HI 96720; westm@hawaii.edu}


\begin{abstract}
The Advanced Camera for Surveys (ACS) Virgo Cluster Survey is
a large program to image 100 early-type Virgo galaxies using 
the F475W and F850LP bandpasses of the Wide Field Channel of 
the ACS instrument on the {\it Hubble Space Telescope} ({\it HST}).
The scientific goals of this survey include an exploration of the
three-dimensional structure of the Virgo Cluster and a critical
examination of the usefulness of the globular cluster luminosity
function as a distance indicator. Both of these issues require
accurate distances for the full sample of 100 program galaxies.
In this paper, we describe our data reduction procedures and
examine the feasibility of accurate distance measurements
using the method of surface brightness fluctuations (SBF) applied
to the ACS Virgo Cluster Survey F850LP imaging.
The ACS exhibits significant geometrical distortions due to its
off-axis location in the {\it HST} focal plane; correcting for these
distortions by resampling the pixel values onto an undistorted frame
results in pixel correlations that depend on the nature of the
interpolation kernel used for the resampling.  This poses a major
challenge for the SBF technique, which normally assumes a flat power
spectrum for the noise.  We investigate a number of different
interpolation kernels and show through an analysis of simulated galaxy
images having realistic noise properties that it is possible,
depending on the kernel, to measure SBF distances using
distortion-corrected ACS images without introducing significant
additional error from the resampling.  We conclude by showing examples
of real image power spectra from our survey.
\end{abstract}


\section{Introduction}

The Virgo Cluster is an appealing target for the study of
early-type galaxy formation and 
evolution because of its richness and its proximity. 
The installation of the Advanced Camera for Surveys 
(ACS; Ford et al. 1998) on the Hubble 
Space Telescope (\hst) has provided the opportunity to observe Virgo galaxies with
unprecedented spatial resolution in reasonable exposure times.
The ACS Virgo Cluster Survey is a program 
to image 100 early-type members of Virgo in the F475W ($\approx$ SDSS {g})
and F850LP ($\approx$ SDSS {z}) filters of the ACS Wide Field Channel (WFC).
This survey aims to study the properties of the globular 
cluster systems associated with the program galaxies, examine
the nuclear properties and the morphological structure of their 
central regions and explore the three--dimensional
structure of the cluster using surface brightness fluctuations (SBF).
A comprehensive description of the program is given in C\^ot\'e et al. 
(2004; hereafter Paper~I). In this paper, we discuss the feasibility
of SBF measurements with the ACS. 

Over the last ten years, the SBF method has been successfully
used to measure early--type galaxy distances out to $\sim\,$7000~km/s
from ground--based telescopes and \hst\
(\citealp{lup93,pah94};
Sodemann \& Thomsen 1995, 1996;
Ajhar \etal\ 1997, 2001;
\citealp{tho97};
Tonry \etal\ 1997, 2001; Lauer \etal\ 1998;
Jensen \etal\ 1998, 1999, 2001;
Blakeslee \etal\ 1999b, 2001, 2002;
\citealp{pah99};
Mei \etal\ 2000, 2001a,c, 2003;
Liu \etal\ 2001, 2002;
Mieske \& Hilker 2003; Mieske \etal\ 2003;
Jerjen \etal\ 2001, 2003, 2004).
%
The method was introduced by Tonry \& Schneider (1988)
and is based on the fact 
that the Poissonian distribution of unresolved stars in a galaxy 
produces fluctuations in each pixel of the galaxy image (see, e.g.,
the review by Blakeslee et al. 1999a).
The variance of these fluctuations is inversely proportional to the square of the galaxy distance and depends linearly on the mean flux of the galaxy. 
To make these fluctuations constant on the galaxy image, the SBF amplitude is defined as this variance normalized to the mean flux of the galaxy in each 
pixel \cite{ts88}. The absolute magnitude of the fluctuations depends on the age and
metallicity of the stellar populations in the galaxy, which means that,
in practice, the calibration of the SBF absolute magnitude for distance
measurements depends on both the observational bandpass and the galaxy color.
 
Different groups have quantified these
dependencies for ground--based and \hst\ photometric bands,
empirically calibrating the dependence of the SBF magnitude on galaxy color  
(\citealp{ton91,ton97,aj97}; Jensen \etal\ 1998, 2003; \citealp{fer00a,bla01b,liu02}).
The dependence of SBF magnitude on color has also been
studied extensively from a theoretical perspective using stellar
population models (Worthey 1993; Buzzoni 1993; Liu \etal\ 2000;
Blakeslee \etal\ 2001; Cantiello \etal\ 2003).
At present, the absolute calibration is based on a comparison of SBF
distances for galaxies with measured Cepheid distances.

In using the ACS for SBF distance measurements, we are faced with two
main obstacles.  First, the ACS cameras suffer from significant geometrical
distortions (Meurer et al. 2002).  Correcting for these distortions involves resampling the pixels
onto a rectified output grid which, depending on the choice of interpolation
used for the resampling, may result in strong noise correlations that could
bias the Fourier-space SBF measurements.  It is therefore important to
characterize this bias in order to ensure an accurate measurement of the
Poissonian stellar population fluctuations.
The second concern involves the proper calibration of the SBF method
for the ACS passbands: a prerequisite for the measurement of SBF distances.
Our survey will provide the first calibration
of SBF distances in the ACS F850LP filter, which, as discussed in Paper~I,
is expected to be nearly optimal for the SBF technique.
The first point will be addressed here, while the ACS SBF calibration will be
the subject of a future paper. 

The plan of this paper is as follows. 
The following section discusses the ACS observations and the SBF data reduction
procedures. In \S~3, we investigate different interpolation kernels and
demonstrate with simulations that our data reduction
procedures allow us to make high-quality, essentially unbiased,
SBF measurements. We conclude in~\S~4.

\section{Surface Brightness Fluctuation measurements with the ACS}

A comprehensive description of the ACS Virgo Cluster Survey observations and
initial data reduction can be found in Paper~I and in Jord\'an et al. (2004;
hereafter Paper~II).  We summarize here the main points.

\subsection{Observations and data reduction}

The ACS Virgo Cluster Survey consists of two-filter (F475W and F850LP) imaging of 100 early-type galaxies 
in the Virgo Cluster,
each one observed within a single HST orbit using the ACS/WFC.
The data for each galaxy consists of five images: two exposure of 375~sec in the F475W filter 
and three exposures in the F850LP filter (two 560~sec and one 90~sec exposures,
 for a total exposure time of 1210~sec).

The ACS detectors have strong geometrical distortions caused mainly by the 
off--axis location of the instrument \cite{meu02}. This distortion causes the 
square pixels of the detectors to project to trapezoids with an area that
varies across the field of view.
By modeling and calibrating these distortions, it is possible to produce a 
geometrically correct image having a constant pixel area.  
As described in Paper~II, we have used the {\it drizzle} software 
(Fruchter \& Hook 2002) to combine the individual exposures of a given
field in our program into a single geometrically corrected image for
each filter. 
However, the interpolation necessary to generate this image can introduce
correlations between output pixels.
Since SBF is calculated as the variance of the Poissonian distribution of
stars in each pixel, SBF measurements can be biased by this correlation.
The amount of correlation that is introduced between nearby pixels
depends sensitively on the choice of the {\it drizzle} interpolation kernel. 
If the images are not corrected for the distortion, and then not interpolated,
the pixel area varies over the image (due to the distortion), 
also potentially biasing the measurement of SBF, 
since Poissonian fluctuations would be measured in pixels of different areas.
Further, if there is dithering between multiple exposures,
distortion correction is necessary in order to combine the images.
We therefore measure SBF in corrected images, quantifying the
bias in the SBF measurements by simulations. 
We describe the SBF reduction in the next section and discuss 
how we can obtain high-quality SBF measurements despite 
the geometrical distortion correction.

\subsection{SBF measurements}
\label{sec:sbfmeas}
We measure the SBF signal in the F850LP data using the standard
SBF extraction techniques (e.g., Tonry \& Schneider 1988; Tonry et al.\ 1991).
The SBF in the F475W data is too faint to measure accurately, but the
galaxy color information from the two bandpasses will be an essential
part of the ACS SBF calibration.
The image processing that we perform in order to produce uniform, galaxy-subtracted
images has been described in more detail by Paper~II; here we summarize the 
main steps.
We fit a smooth model for each galaxy in both filters using ELLIPROF (the isophotal
fitting software that has been used for the SBF survey by Tonry et al. 1997) and
then subtract the models from the original images. To eliminate
 large-scale residuals left from this subtraction, we run SExtractor and produce a 
smooth background model which we then subtract from the 
galaxy-subtracted image.
Object detection is then performed on this residual image, using a 
detection threshold of five connected pixels at 1.5~$\sigma$ significance
level (see Paper~II for the details of these first steps of the data reduction).

A catalog of sources is produced by matching the F475W and F850LP detections.
This procedure is necessary to separate
spurious detections (for instance, residual cosmic ray events or hot
pixels) from real, faint objects. Red background galaxies, however,
might produce a clear detection in F850LP only; we therefore added to
the list of sources detected in both filters objects brighter than 23~mag 
in F850LP, but absent in the F475W frames. The source catalog
was then trimmed by excluding objects fainter than a cut-off magnitude
$m_{cut}$. 

The final sources list
thus obtained was used to create a ``source mask,'' to remove
the contribution to the image power spectrum from contaminating sources
such as foreground stars, globular clusters, and background galaxies.
We mask each source using an aperture with size equal to the
maximum of 3 times the FWHM (Full Width Half Maximum) 
of the PSF (point spread function) 
(usually best adapted for compact objects) 
and the product of the Kron radius and the object major axis 
(best adapted to extended objects) as calculated by SExtractor.

Detected sources  with magnitude brighter than the cut--off
magnitude $m_{cut}$ 
are removed from the residual image by multiplying it by the source mask.
The resulting image is then divided into concentric annuli centered on the
galaxy center, and the power spectrum of each annulus is measured.
To do this, we multiply the galaxy image by a total annular mask function 
defined as the product of the source 
mask function and the annular mask function.
Assuming no noise correlations, 
the image power spectrum is the sum of two simple and distinct 
components: the flat, white noise power spectrum and the combined power
spectrum of the SBF and the undetected faint sources, 
both convolved by the PSF in the spatial domain.

In the Fourier domain, the convolution by the PSF translates to 
a multiplication.  We therefore fitted the azimuthally averaged 
image power spectrum $P(k)$ to the function
\begin{equation}
P(k) = P_0 \times E(k) + P_1 \,,
\label{eq:pk}
\end{equation}
where the ``expectation power spectrum'' $E(k)$ is  the
convolution of the power spectra of the normalized PSF and of the
mask function of the annular region being analyzed. 
In equation~\ref{eq:pk}, $P_0$
represents the PSF-convolved component, while $P_1$ is the white noise
component which is modified by the distortion correction (see below).
A high signal-to-noise composite $z$-band PSF was created using
NGC\,104 (47~Tuc) observations from \hst\ calibration program GO-9018.
After normalizing the total flux to 1~electron per second, its power
spectrum was computed.
While the PSF of the ACS WFC does show some spatial variation \cite{kri03},
the variation in the $z$-band is not significant for our purposes
as it is present on very small scales that do not significantly
affect our power spectrum fits (see below).
To verify this, we investigated the effects of using individual stars as
PSF templates, and found that the typical variation in the fitted $P_0$ was
$\approx$ 2\%.

We wish to normalize the $P_0$ component by the mean galaxy surface brightness
in the region under consideration so that the SBF contribution to $P_0$
is equal to the ratio of the second and first moments of the stellar
luminosity function (e.g., Tonry \& Schneider 1988).
This is done by multiplying $E(k)$, as defined above, by the square
root of the galaxy surface brightness model for the annulus (Tonry et
al. 1991). In practice, this is achieved by defining  a ``window
function'' as:
\begin{equation}
W({\bf x}) = M ({\bf x}) \times \sqrt{G({\bf x})}
\end{equation}
where $M ({\bf x})$ is the source mask and $G({\bf x})$
is the galaxy surface brightness model for the annulus.
The 2-D expectation power spectrum is then
\begin{equation}
E({\bf k}) = P_{PSF}({\bf k}) \otimes P_{W}({\bf k})
\end{equation}
where $P_{PSF}({\bf k})$ is the spectrum of the normalized PSF and
$P_{W}({\bf k})$ is the power spectrum of the window function.  
The power spectrum is then azimuthally averaged, and a
linear fit to equation~(\ref{eq:pk}) is made by 
minimizing the absolute
deviation (Press et al. 1992) to derive $P_0$ and $P_1$. 
The uncertainty in the fit, including variations resulting from 
different choices for the adopted $k$ range, is included in our
final error estimates. A complete description of the errors will be
given in the ACS SBF calibration paper.

To our knowledge, all previous SBF measurements have been made
without ever resampling the image pixel values (e.g., not correcting
for distortions and allowing only integer pixel shifts);  however, 
the instruments used for past SBF observations had significantly
less distortion than the ACS.
The measurement of $P_0$ and $P_1$ in distortion-corrected images
can be biased by the interpolation
kernel used for the pixel resampling since the noise
correlations (introduced by interpolation) change the shape of the
power spectrum at the correlation scale.  The choice of the
interpolation kernel is therefore a critical aspect of the ACS data reduction,
to which we now turn our attention.

\section{Simulations of power spectrum biases and results}

We have studied in detail the effects of four possible choices of the {\it drizzle}
interpolation kernel on the SBF measurements: the {\it square} (Bilinear), {\it gaussian} (Gaussian),
{\it lanczos3} (Lancsoz3; a damped sinc function), and {\it point} (Nearest pixel)
interpolation kernels.
The sinc function is the ideal low-pass filter, but it cannot be used by itself due
to its asymptotically oscillating character. When used as an interpolation filter, it
is usually multiplied by a window function.
The 3-lobed Lancsoz-windowed sinc function that we use is defined as
\begin{eqnarray}
Lancsoz3(x) = \left \{ \begin{array}{l}
\frac{sin(\pi x )}{\pi x}  \frac{sin(\pi x/3 )}{\pi x/3} \hspace{1cm} 
|x|<3 \\
{0       \hspace{3.3cm}    |x|>3 } \end{array}
\right.
\end{eqnarray}

To examine the effects of the interpolation kernel on the SBF
measurements, we have generated two sets of simulated images. The first
set contained simply a white noise component, while the second set
consisted of model galaxies having total luminosities representative
of the galaxies in the ACS Virgo Cluster Survey sample. 
The goal of these simulations is to study the effect of the 
pixel resampling under different interpolation kernels during 
drizzling, rather than to simulate the ACS Virgo galaxy observations
in full detail.  Therefore, the simulated raw images do not
have distorted pixels, but both sets of simulations are geometrically
transformed using {\it drizzle} in the same way that the actual distorted
images are corrected in our ACS pipeline.

\subsection{Results from white noise simulations}

\begin{figure}
\epsscale{.80}
\centerline{\plotone{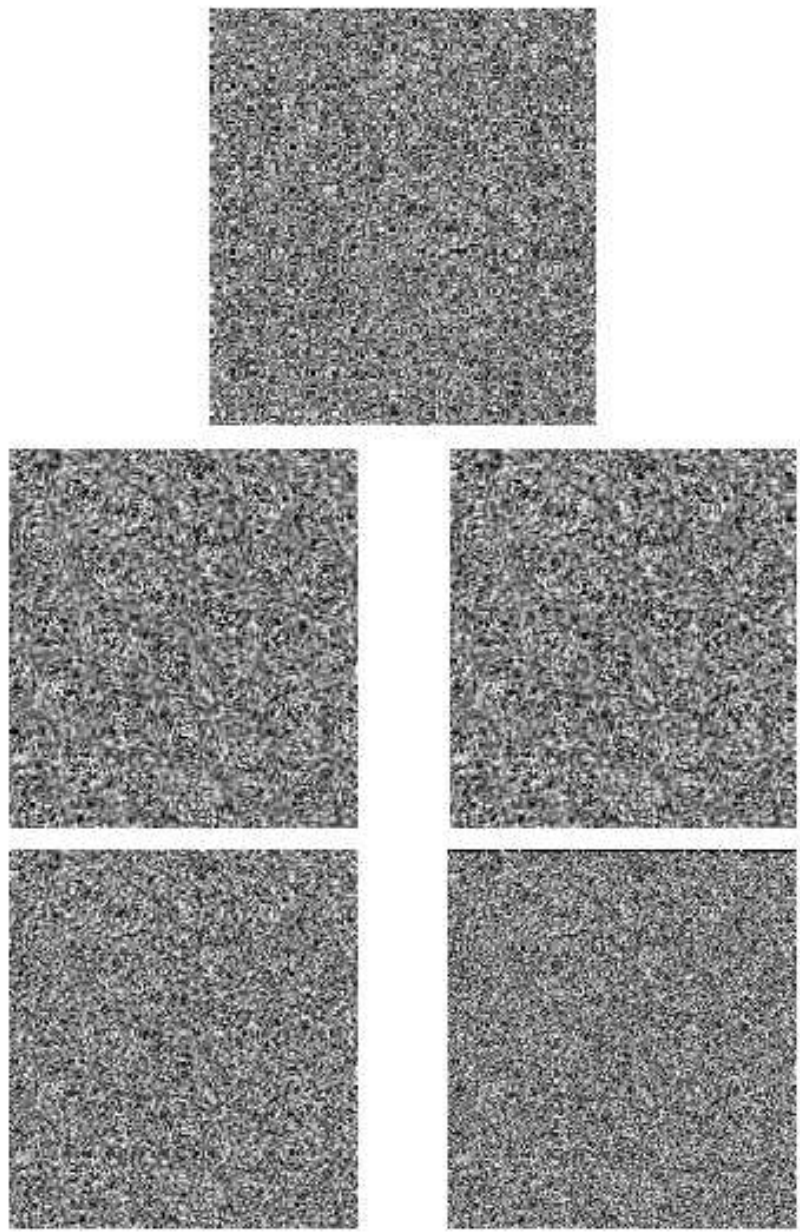}}
\caption{Results from ``drizzling'' white noise simulations
with different interpolation kernels.
On the very top at the center is shown the white noise simulation before drizzling.
The central two images have been drizzled with the Bilinear (left) and 
Gaussian (right) kernels.
The lower images have beens drizzled with the Lanczos3 (left) and Nearest 
(right) kernels.\label{images}}
\end{figure}

Sample images for the white noise simulations are shown in Fig.~\ref{images}.
It is apparent that both the Bilinear and Gaussian images
show visible patterns of correlation.
The Nearest and Lanczos3 kernels introduce less correlation. 
The Nearest interpolation leaves a large
number of zero (uninterpolated) pixels in the output 
image and does not preserve
relative astrometry at the pixel scale.
This is because no real interpolation is done in this case; 
the input pixel value is simply assigned to the nearest output
pixel.
 
\begin{figure}
\epsscale{.80}
\plotone{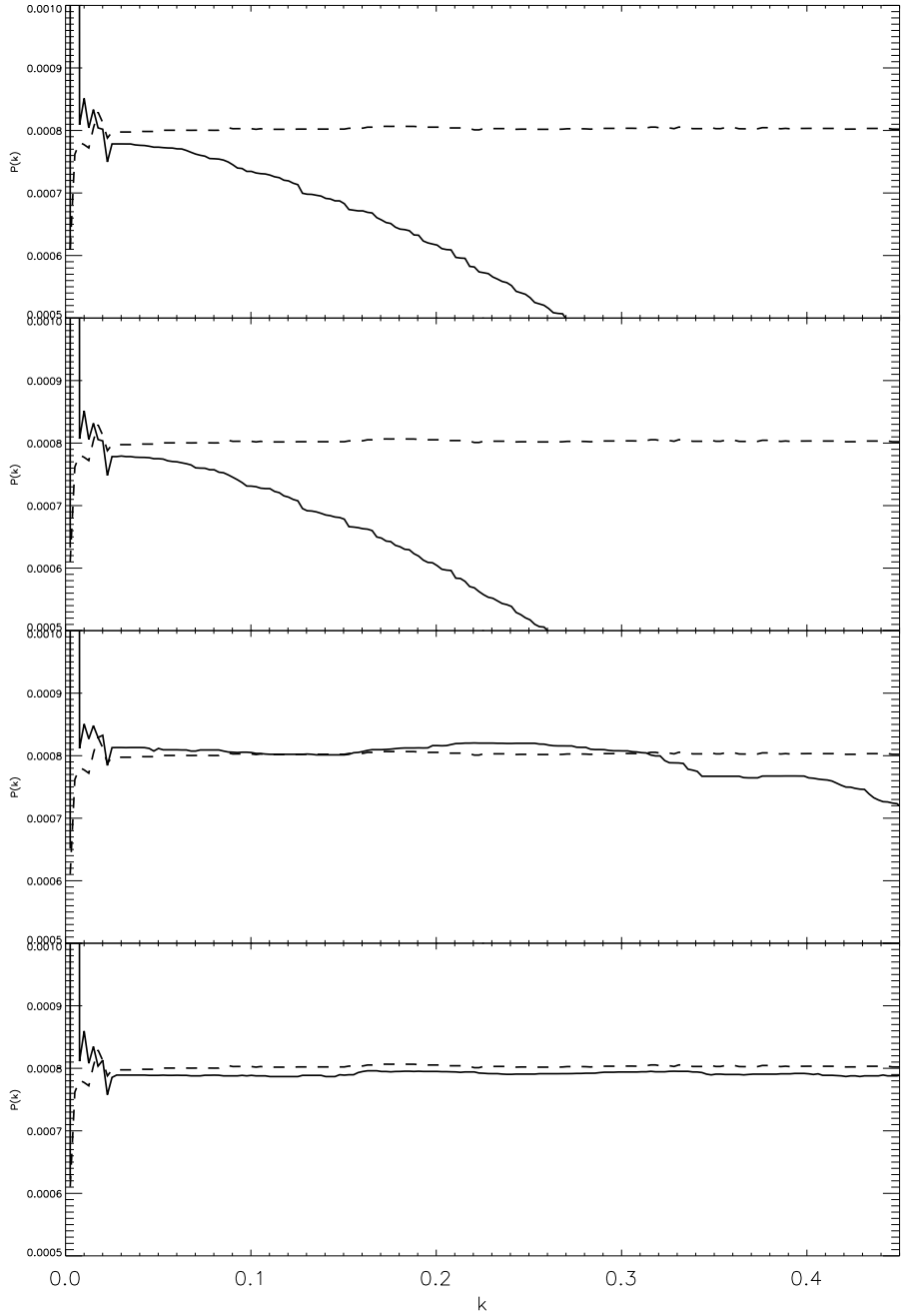}
\caption{Noise power spectra when different interpolation kernels 
are used in the image drizzling. From top to bottom, the four panels
refer to the Bilinear, Gaussian, Lanczos3, and Nearest kernels.
The wavenumber $k$ corresponds to the inverse of the pixel scale.
The continuous lines are the spectra after drizzling, the dashed line 
is the power spectra of white noise before drizzling.
\label{fig1}}
\end{figure}

\begin{figure}
\epsscale{.80}
\plotone{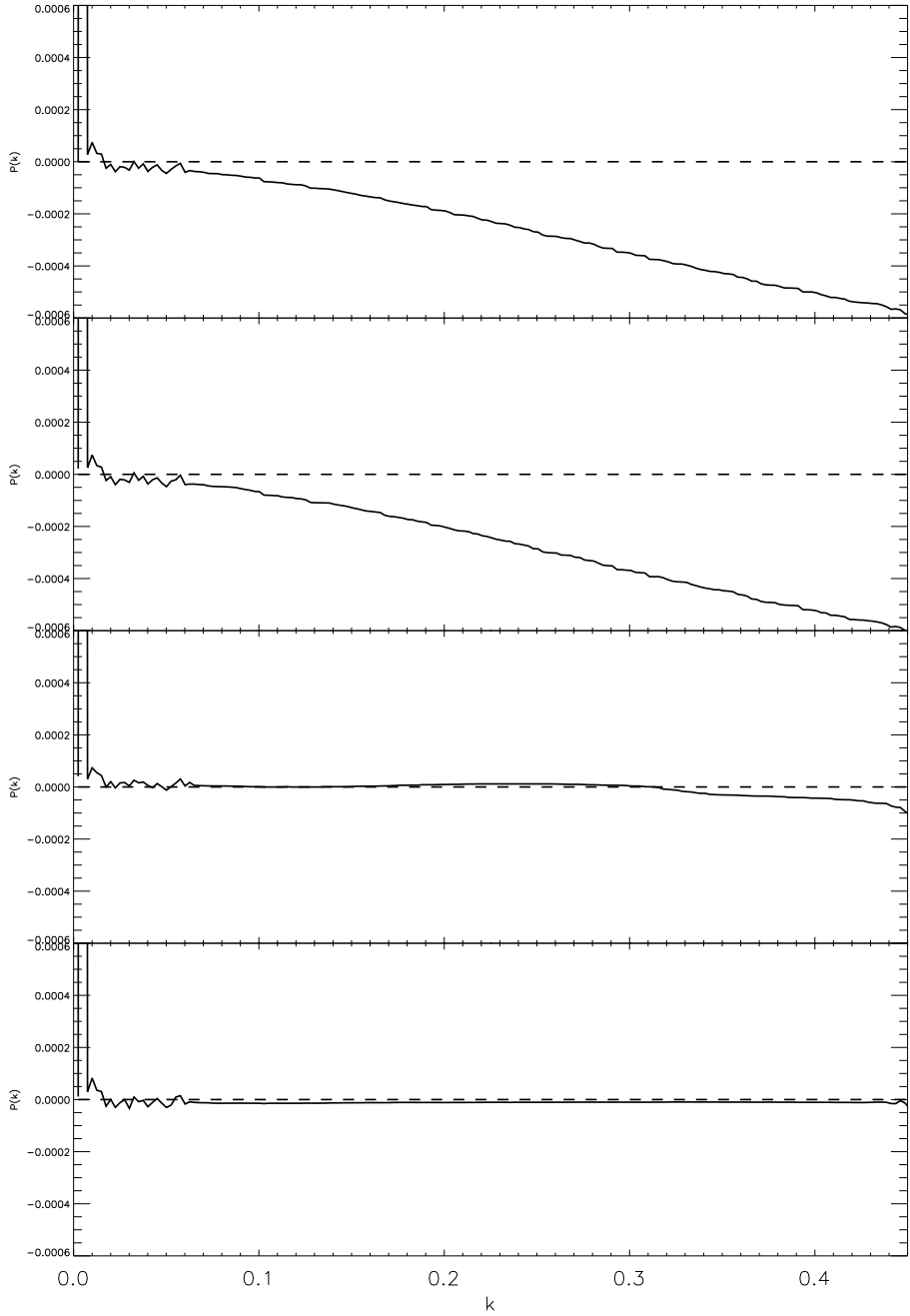}
\caption{Differences between the original and drizzled noise
power spectra are shown as continuous lines for the same four 
cases as in Fig.~\ref{fig1}. From
top to bottom, the panels refer to the Bilinear, Gaussian, Lanczos3,
and Nearest kernels. The dashed line is the case where the difference
is zero.
\label{fig2}}
\end{figure}

To quantify the bias arising from the interpolation, we have calculated
power spectra for the simulated images. For the white noise simulation,
the original image has, as expected, a flat power spectrum.  However, once
the drizzling has been applied, both low and high wavenumbers are
contaminated by the pixel correlations introduced by the
interpolation kernels.  We show in Fig.~\ref{fig1} our results for the
four different kernels. In all cases, the original power spectrum is shown
as a dashed line while the power spectrum of the drizzled image is shown
as the continuous line. For comparison, 
Fig.~\ref{fig2} shows the difference between the
power spectra measured in the original and drizzled images.

As Fig.~\ref{fig1}\,--\,\ref{fig2} reveal, the Bilinear and Gaussian interpolations
introduce artificial pixel correlation
on all scales, making it impossible to use these kernels for accurate SBF measurements.
The Lanczos3 kernel depresses the power spectrum at high wavenumber (small pixel scales).
The Nearest kernel does not bias in any
significant way the image power spectrum, which
remains flat (constant with k), even if the power spectrum is
systematically lower than the original.
From the noise simulations, both of the latter kernels can be used for SBF 
measurements in the k range in which the power spectrum remains flat ( i.e. 
provided the smallest pixel scales
are not used when fitting the power spectrum with the Lanczos3 interpolation).
From the white noise simulations drizzled with the Lanczos3 kernel,
we have measured the pixel scale range in which the power spectrum 
is flat to within 3~$\sigma$. This is true for scales between 3 
and 20 pixels, corresponding to fractional wavenumbers in the range
0.05 $\le k \le$ 0.33.  On smaller pixel scales, the original white noise power 
spectrum is significantly depressed by the noise correlation.

\subsection{Results from the galaxy simulations}

The galaxy simulations have
been performed following the method described in Mei et al. (2001b).
With the galaxy magnitude and effective radius specified, 
the surface brightness at each pixel is calculated using a 
de\,Vaucouleurs (1948) profile and Bruzual \& Charlot (2003) stellar
population models corresponding to solar metallicity and an age of 12 Gyr. 
The results are not sensitive to the use of the de\,Vaucouleurs profile,
since the SBF measurements are done on the galaxy model-subtracted images.
The isophotal model used for the galaxy subtraction does not assume 
the de\,Vaucouleurs form (see \S\,\ref{sec:sbfmeas}).
Just as in real data, mismatches between
the simulated galaxy images and isophotal models will produce
large-scale spatial signatures in the power spectra that
do not affect our fitting scales.  Similarly, the precise age and
metallicity of the stellar population model is not important, 
except insofar as they precisely specify the SBF magnitude 
for the simulations.

\begin{figure}
\epsscale{.80}
\centerline{\plotone{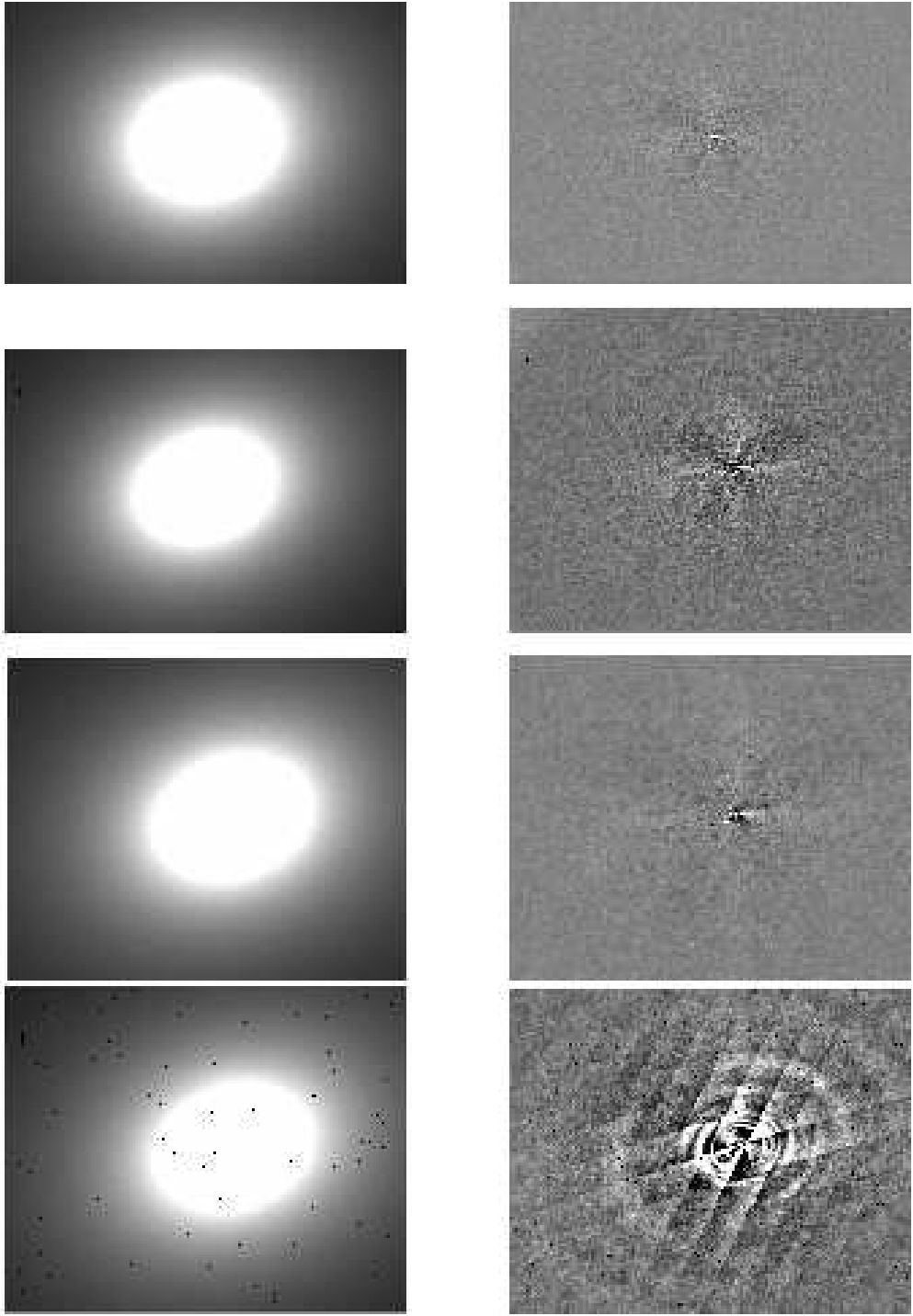}}

\caption{Results from  galaxy simulations
with different interpolation kernels.
On the very top is shown a simulation (before drizzling) for a galaxy 
($\approx 30~\arcsec \times 20~\arcsec$) with $M_B = -22$ (left), and the residual
after subtracting a smooth galaxy model (right). 
The residuals show fluctuations from Poissonian noise, read out 
noise and the PSF-convolved Poissonian stellar fluctuations.
In the center images, the same for the same galaxy image drizzled 
with a Gaussian and a Lancoz3 filter.
In the bottom, the same, drizzled with a Nearest filter.
\label{galaxy_image}}
\end{figure}

\begin{figure}
\epsscale{.80}
\centerline{\plotone{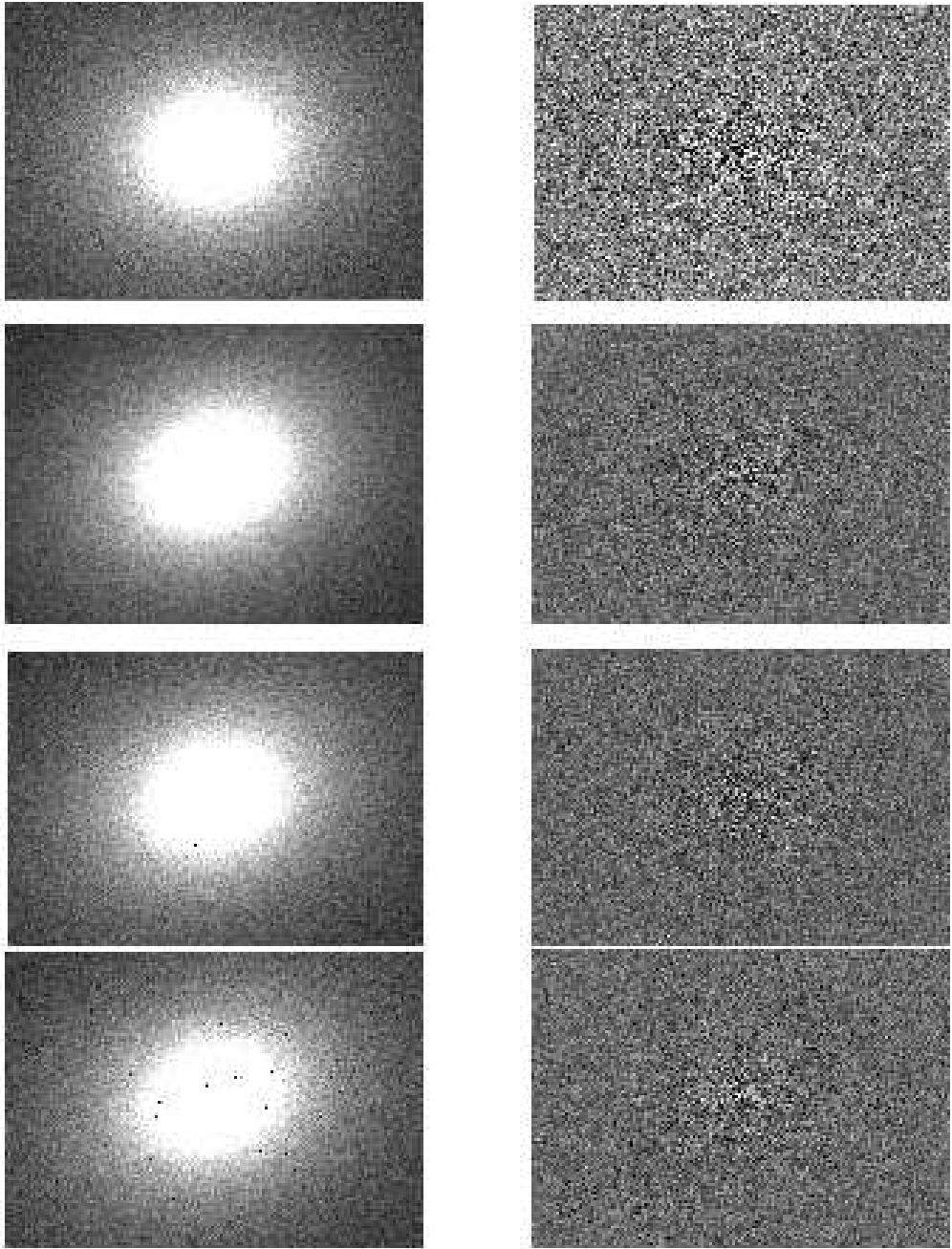}}

\caption{Results from  galaxy simulations
with different interpolation kernels.
the same than Fig.~\ref{galaxy_image}, 
for a galaxy 
($\approx 30~\arcsec \times 20~\arcsec$) with $M_B = -17$.
\label{galaxy_image2}}
\end{figure}

To each pixel in the simulated images,
we add a Poisson variance calculated from the number of stars 
contributing to the pixel, Poissonian noise from the galaxy and the sky 
(22 mag\,arcesc$^{-2}$), and a read noise of 7~$e^-$.
The images are then convolved with the PSF and drizzled using
three different interpolation kernels: Gaussian, Lancsoz3 and Nearest. 
Since the results from the white noise simulations show that the Gaussian
and Bilinear kernels introduce similar strong pixel correlations,
we use only the Gaussian to quantify this effect.
Moreover, in the larger context of our survey, this kernel allows
a more faithful reconstruction of the galaxy centers.
Galaxy simulation images are shown in  Fig.~\ref{galaxy_image} and  Fig.~\ref{galaxy_image2}.
The left panels show the simulated galaxies and the right panels
show the residuals following subtraction of smooth isophotal models.
As for the white noise simulations, the Gaussian interpolation
shows visible patterns in all the residuals.
The Nearest interpolation shows visible residual patterns in 
the centers of the galaxy simulations where the intensity profile is steepest,
especially for the brighter galaxy.

To quantify the effect of the interpolation on the SBF
measurements, we have simulated 100 galaxies for an assumed Virgo distance
of 16~Mpc, and for three galaxy magnitudes corresponding to:
 1) the brightest galaxy in our
sample ($M_B = -$22~mag); 
2) the faintest galaxy ($M_B = -$15~mag);  
and 3) an intermediate galaxy ($M_B = -$17~mag).
The $B-V$ (Vega) color from Bruzual \& Charlot (2003) models is 0.97~mag, and
the $V-F850LP$ (AB) color is 0.9~mag.
Each of these simulations has been drizzled with the three different kernels.
We then applied the SBF reduction procedure, as described above, to
the simulated galaxy images to obtain SBF magnitudes, and 
then distances from the input SBF absolute magnitude 
($\overline M_{F850LP}=-1.42$
from the Bruzual \& Charlot stellar population model).

\begin{figure}
\epsscale{0.9}
\plotone{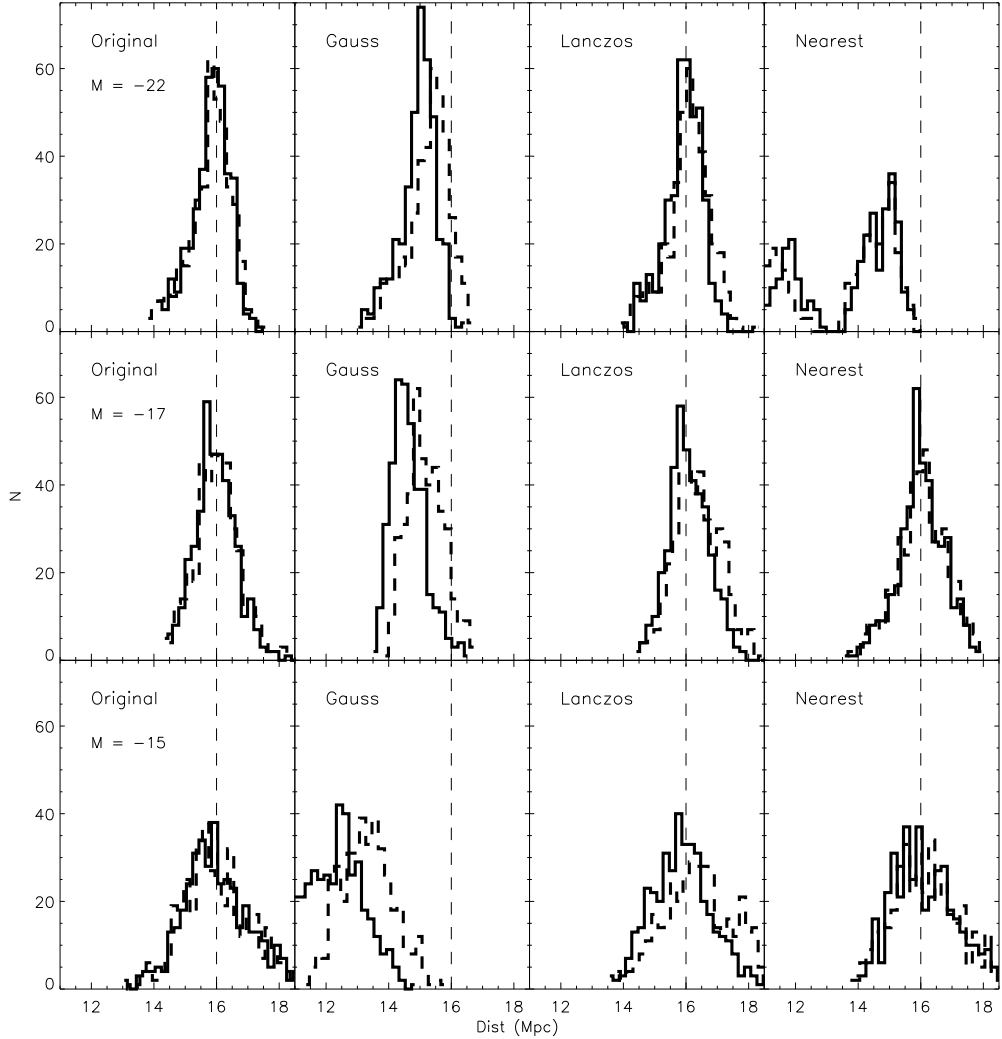}
\caption{The histograms of the measured SBF distances are shown
for the three different magnitude simulations. For each of the
100 simulations at each magnitude, SBF magnitudes and distances
have been measured in four different annuli having inner and outer radii of 
1\farcs6 and 3\farcs2, 3\farcs2 and 4\farcs8,  4\farcs8 and 6\farcs4,
 and 6\farcs4 and 8\arcsec.
From left to right we show results for the original 
galaxy simulations, and, respectively, for the Gaussian, Lanczos3, and 
Point/Nearest drizzled images.
From top to bottom the galaxy absolute B magnitude is $-$22~mag, $-$17~mag
and $-$15~mag. The solid histograms correspond to results obtained with a
fit on the entire scale range, the dashed ones to fits obtained in the range
between 3 to 20 pixels. The dashed
 vertical line corresponds to the input distance of 16~Mpc.  
\label{sbfh}}
\end{figure}

Histograms of the resulting SBF distances for individual annuli of
the 100 galaxy image realizations are shown in Fig.~\ref{sbfh}
and summarized in Table~\ref{results}.
SBF magnitudes and distances are shown when the fit is performed
in two different ranges:  (1) between 3 to 20 pixels, and 
(2) between 0 to 20 pixels, i.e., out to the maximum wavenumber. 
The fit was not extended beyond 20 pixels because, in the real data,
larger scales will be biased by the subtraction of the large--scale
residuals (see \S~2.2).
 
The Gaussian filter introduces a bias in SBF distance measurements
of 6 to 20\%, compared to a bias of 0.1 to 4\% for  the Lancsoz3 kernel.
The Nearest filter tends to underestimate the distance in the innermost annuli
of bright galaxies. As seen in Fig.~\ref{sbfh}, when the Nearest filter is used,
the distance histogram of 
the brightest galaxy shows a bimodal distribution, 
with peaks at $\approx 11$~Mpc and $\approx 15$~Mpc,
corresponding to the innermost and outermost annuli,  respectively.
This causes a bias of 20\% on the total SBF distance of bright galaxies.
For fainter galaxies, the biases due to this filter are between 1 and 2 \%.
However, contrary to expectations from the white noise simulations,
when the Lanczos3 filter is used, galaxy distances do not significantly 
improve if we use a maximum wavenumber corresponding to 3 pixels.
This is because of the reduced leverage for determining the $P_1$
value in equation~(\ref{eq:pk}).

\begin{figure}
\plotone{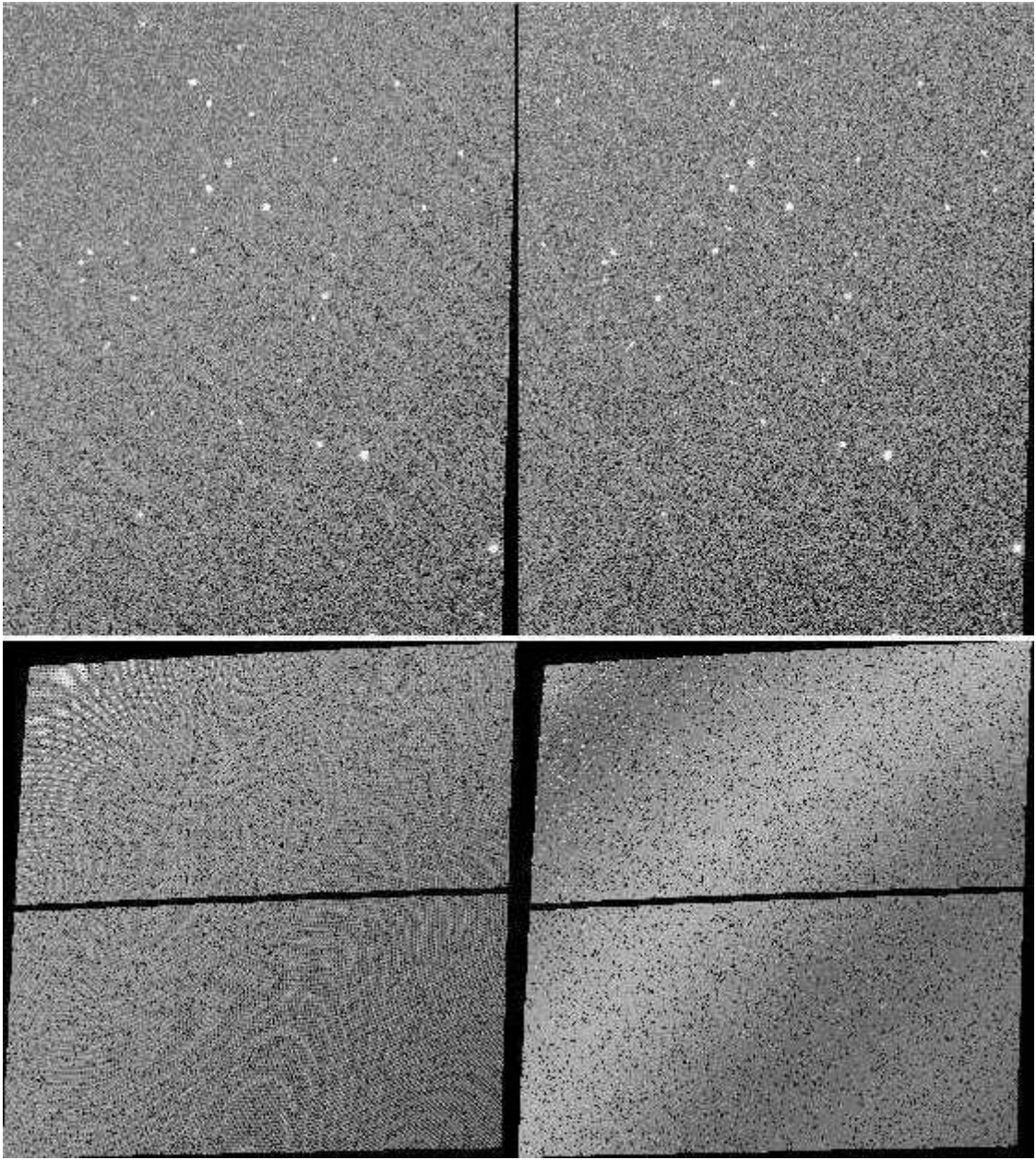}
\caption{Top: a $0\farcm5{\,\times\,}0\farcm7$ section of the M87
image is shown when the Gaussian (left) and Lanczos3 (right) drizzle
interpolation kernels are used. Bottom: the full {\it drizzle}
weight images obtained for these interpolations (including weight
variations from rejection of pixels affected by cosmic rays and
detector defects), again with the Gaussian results on the left
and the Lanczos3 results are on the right.
\label{kernels}}
\end{figure}

We have chosen to use the Lanczos3 kernel for our final SBF analysis
to insure a homogeneous data treatment between faint and bright galaxies,
and because it preserves astrometric fidelity.  
This has the further advantage that
the same images can be used for fitting globular cluster structural parameters.
However, since the  Lanczos3 kernel is not well adapted to isophotal analysis
(e.g., negative pixel values occur near saturated or repaired pixels),
we have produced a second set of images with a Gaussian kernel, used
in the isophotal analysis of the program galaxies.
As a visual comparison of the effects on real data, 
the top panel of Fig.~\ref{kernels} 
shows a
corner of the M87 image drizzled with the Gaussian (left) and the
Lanczos3 (right) kernel.  The lower panel compares the respective
weight images for the two kernels as derived from the {\it drizzle}
interpolation (see Paper~II).

\begin{figure}
\epsscale{1}
\plottwo{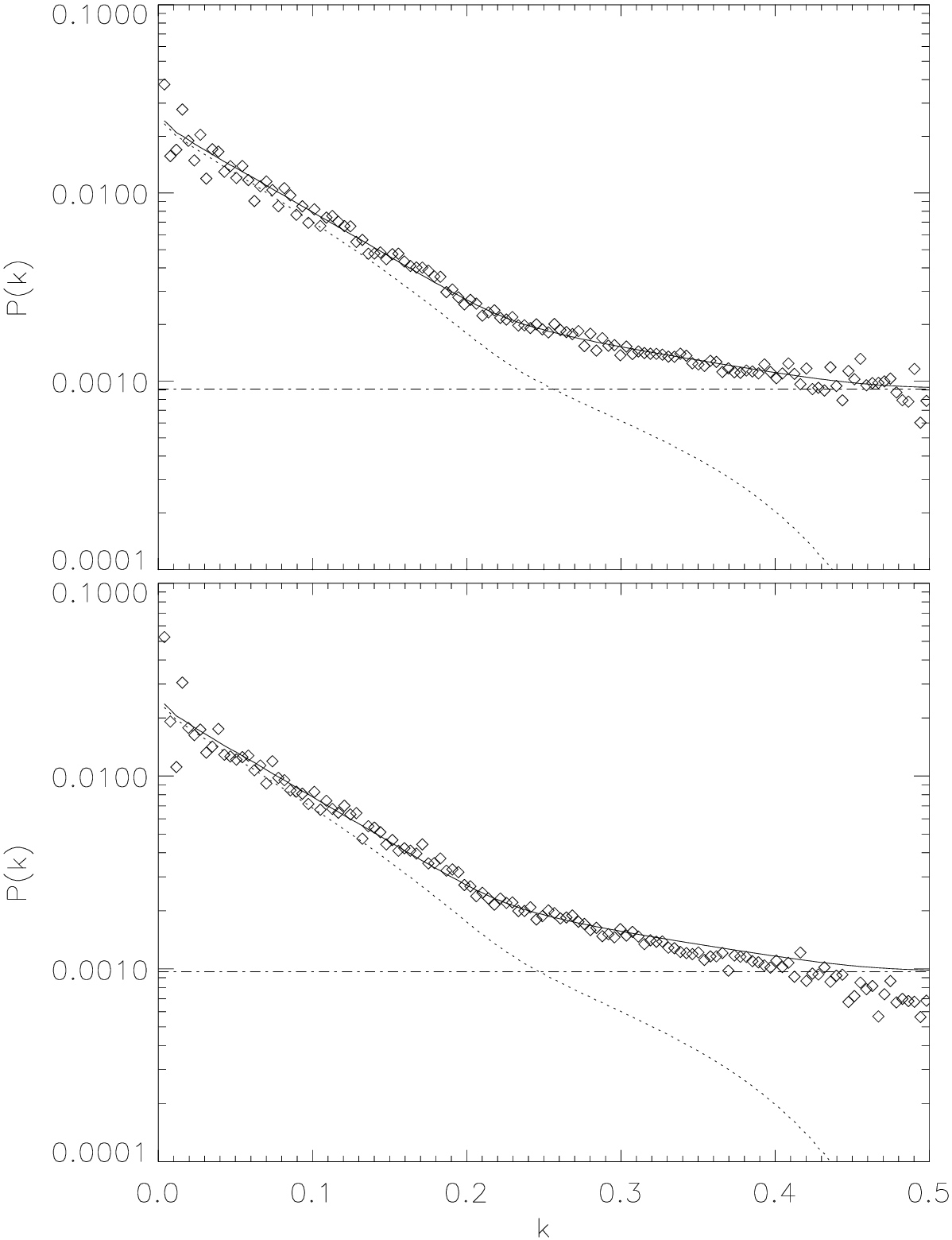}{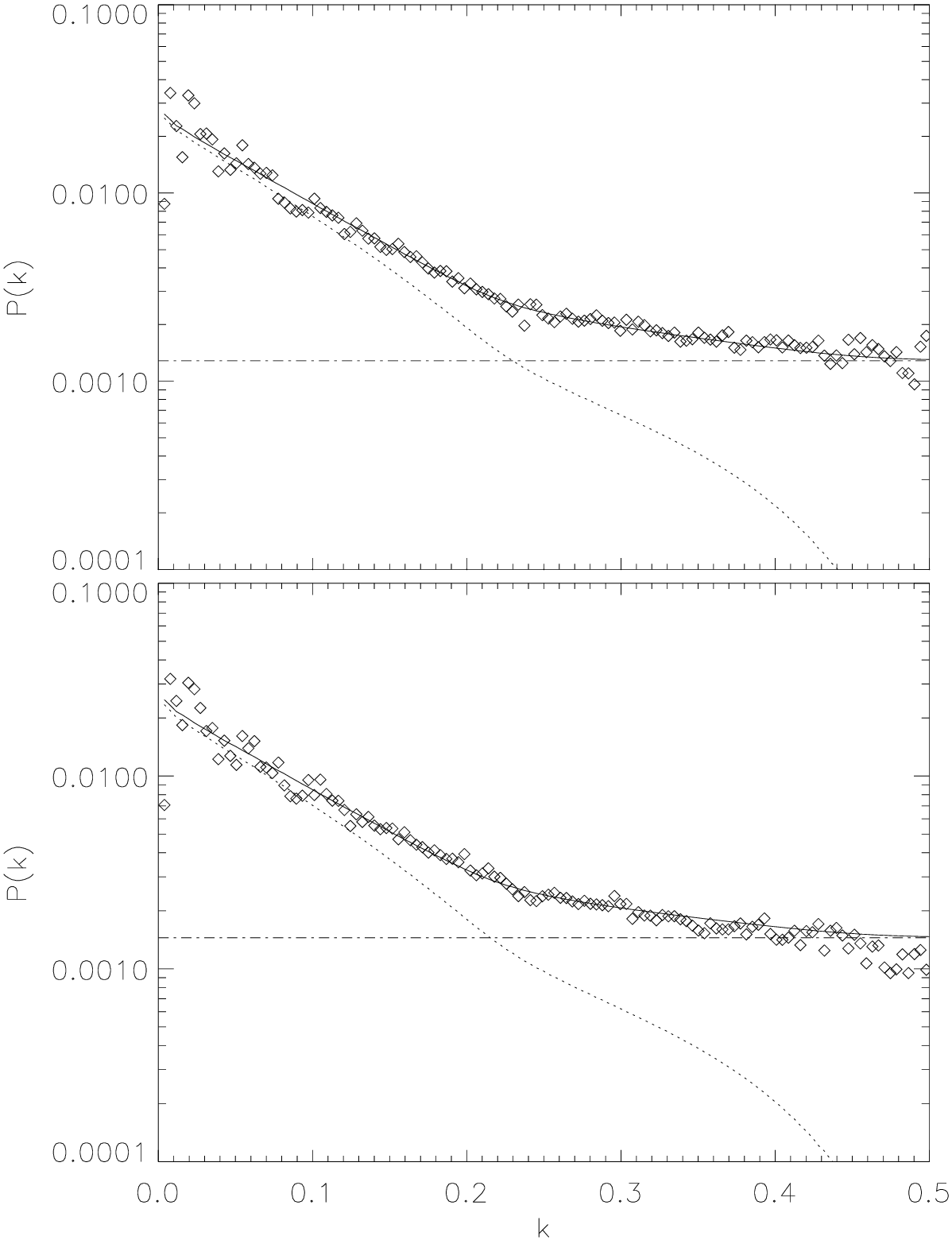}
\caption{Original (top) and drizzled (bottom) power spectra
for the galaxy simulations when a Lanczos3 kernel is used.  The
results are for an annulus with inner and outer radius of respectively 
$\approx$1 and $\approx$6 \arcsec.  On the left, the spectrum of a galaxy 
with  $M_B = -$22~mag is shown before (top) 
and after (bottom) drizzling; on the right, the same thing is shown for a 
galaxy with absolute $M_B = -$17~mag.
The diamonds are the image power spectra, the continuous line the 
fit of the power spectrum P(k), the dotted line is $P_0 \times E(k)$
and the dot-dashed line indicates the fitted $P_1$.  \label{fig3}}
\end{figure}

\begin{figure}
\epsscale{.70}
\plotone{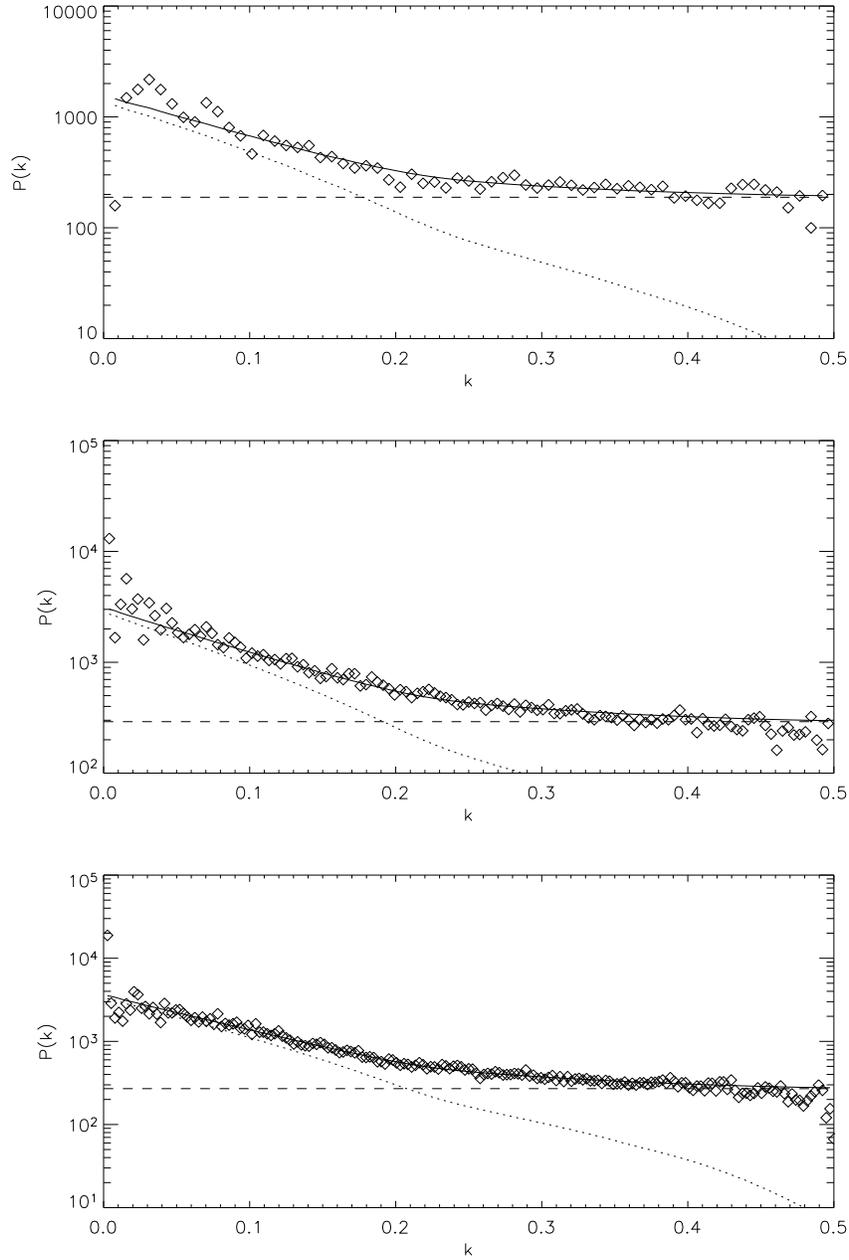}
\caption{VCC1226 (M49, NGC~4472) power spectra for the three innermost
annuli. From top to bottom, the power spectra were calculated using 
annuli having inner and outer radii of 1\farcs6 and 3\farcs2,
3\farcs2 and 6\farcs4, and 6\farcs4 and 9\farcs6.
The diamonds are the image power spectrum, the continuous line the 
fit of the power spectrum P(k), the dotted line is  $P_0 \times E(k)$, and 
the dashed line is $P_1$.  \label{NGC4472}}
\end{figure}

Fig.~\ref{fig3} shows simulated galaxy power spectra in an
annulus with inner and outer radius of, respectively, $\approx 1 \arcsec$ 
and $\approx 6\arcsec$. On the left, the spectrum of a galaxy 
with an absolute $B$-band magnitude of $-$22~mag is shown before (top) 
and after (bottom) drizzling; on the right, the same comparison is given for a 
galaxy with absolute $B$-band magnitude of $-$17~mag.
We have fitted the power spectrum of the original image over the entire 
scale range and the power spectrum of the drizzled image for scales 
between 3 and 20 pixel. For this particular choice of annulus, 
the distance values between 
the original and drizzled simulated images agree to $4\%$, 
with a corresponding difference in SBF magnitudes of $\approx$0.05~mag.
In Fig~\ref{NGC4472} we show NGC~4472 power spectrum for three innermost annuli
 with respective inner and outer radius of 1\farcs6 and 3\farcs2,
3\farcs2 and 6\farcs4, and 6\farcs4 and 9\farcs6.

\section{Conclusions}

We have described the SBF data reduction procedures for the ACS Virgo
Cluster Survey, paying close attention to the feasibility of SBF measurements 
with the ACS/WFC instrument. Although the ACS images exhibit strong
geometrical distortions that could bias SBF measurements, we have 
shown with the aid of simulated galaxy images having
realistic noise properties that it is possible to measure accurate
SBF distances by correcting for biases, taking care to mask
contaminants (i.e., globular clusters, foreground stars and
background galaxies), and using
an appropriate (Lanczos3) interpolation kernel. Future papers in this series
will derive a new $z$-band SBF calibration for our sample, 
present the final SBF distances and their uncertainties, 
compare the SBF and globular cluster luminosity function distances, 
and examine the three-dimensional structure of the Virgo cluster.

\acknowledgments
Support for program GO-9401 was provided through a grant from the Space
Telescope Science Institute, which is operated by the Association of 
Universities for Research in Astronomy, Inc., under NASA contract NAS5-26555. 
ACS was developed under NASA contract NAS 5-32865.
S.M. and J.P.B. acknowledge additional support from NASA grant 
NAG5-7697 to the ACS Team.
A.J. acknowledges support provided by the National
Science Foundation through a grant from the Association of Universities
for Research in Astronomy, Inc., under NSF cooperative agreement
AST-9613615, and by Fundaci\'on Andes under project No.C-13442.
P.C. acknowledges support provided by NASA LTSA grant NAG5-11714.
D.M. is supported by NSF grant AST-020631, 
NASA grant NAG5-9046, and grant HST-AR-09519.01-A from STScI. 
M.M. acknowledges support from the Sherman
M. Fairchild foundation. 
M.J.W. acknowledges support through NSF grant AST-0205960.
This research has made use of the NASA/IPAC Extragalactic Database (NED)
which is operated by the Jet Propulsion Laboratory, California Institute
of Technology, under contract with the National Aeronautics and Space Administration. 
We thank Stephane Charlot for providing theoretical
   stellar population models in the ACS filters, and Michele
   Cantiello and Gerhardt Meurer for useful discussions..

\newpage

\begin{table*}
\begin{center}
\caption{Measurements of SBF Magnitudes and Distances from Drizzled Simulated Images.\label{results}}
\vspace{0.25cm}
\begin{tabular}{ccccccccc}
\tableline \tableline\\
$M_B$ & Kernel & Pixel scale range& SBF mag. &$\sigma_{SBF}$& Dist.&$\sigma_{Dist}$&Bias \\
(mag) &&&(mag)&(mag)&(Mpc)&(Mpc)&\% \\ \\
\tableline \\
$-$22&Original&0-20&29.57&0.08&15.9&0.6&0.6\\
&Gauss&0-20&29.46&0.08&15.0&0.6&6\\
&Lanczos&0-20&29.60&0.08&16.0&0.6&0.1\\
&Nearest&0-20&28.91&0.66&12.2&3.2&24\\ \\
$-$17&Original&0-20&29.60&0.08&16.1&0.7&0.5\\
&Gauss&0-20&29.41&0.08&14.7&0.5&8\\
&Lanczos&0-20&29.62&0.09&16.2&0.6&1\\
&Nearest&0-20&29.61&0.10&16.1&0.8&0.7\\ \\
$-$15&Original&0-20&29.61&0.15&16.2&1.1&1\\ 
&Gauss&0-20&29.04&0.16&12.4&0.9&22\\
&Lanczos&0-20&29.58&0.13&16.0&1.0&0.1\\
&Nearest&0-20&29.62&0.14&16.2&1.1&0.2\\ \\

$-$22&Original&3-20&29.58&0.10&15.9&0.7&0.6\\
&Gauss&3-20&29.50&0.10&15.4&0.7&4\\
&Lanczos&3-20&29.60&0.10&16.2&0.7&1\\
&Nearest&3-20&28.84&0.72&11.9&3.4&26\\ \\

$-$17&Original&3-20&29.61&0.10&16.1&0.7&0.8\\
&Gauss&3-20&29.49&0.09&15.3&0.6&5\\
&Lanczos&3-20&29.65&0.10&16.5&0.7&3\\
&Nearest&3-20&29.61&0.10&16.2&0.8&1\\ \\

$-$15&Original&3-20&29.62&0.17&16.2&1.3&1\\
&Gauss&3-20&29.21&0.14&13.4&0.9&16\\
&Lanczos&3-20&29.68&0.17&16.7&1.3&4\\
&Nearest&3-20&29.63&0.16&16.3&1.2&2\\ \\

\tableline \tableline
\end{tabular}
\end{center}
\end{table*}

\clearpage


\newpage

\begin{thebibliography}{}

\bibitem[Ajhar et al. 2001]{ajh01}
 Ajhar E.A., Tonry, J.L.,  Blakeslee, J.P. et al. 2001, ApJ, 559, 584
 

\bibitem[Ajhar et al. 1997]{aj97}
Ajhar E.A., Lauer, T.R., Tonry, J.L. et al., 1997, AJ, 114, 626



\bibitem [Blakeslee et al. 1999a]{bla99a}
Blakeslee, J. P., Ajhar, E. A., Tonry, J. L., 1999a, in Post-Hipparcos
Cosmic Candles, eds. A. Heck \& F. Caputo (Boston: Kluwer), 181

\bibitem [Blakeslee et al. 1999b]{bla99b}
Blakeslee, J. P.,  Davis, M., Tonry, J. L. et al., 1999b, ApJ, 527, L73 

\bibitem [Blakeslee et al. 2002]{bla02}
Blakeslee, J.P., Lucey, J.R., Tonry, J.L. et al. 2002, MNRAS, 330, 443


\bibitem[Blakeslee et al. 2001]{bla01b}
Blakeslee, J.P., Vazdekis, A., \& Ajhar, E.A.\ 2001, \mnras, 320, 193 

\bibitem[Bruzual \& Charlot(2003)]{bru03} Bruzual, G.~\& 
Charlot, S.\ 2003, \mnras, 344, 1000 

\bibitem[Buzzoni 1993]{bu93}
Buzzoni, A. 1993, A\&A, 275, 433 

\bibitem[Cantiello \etal\ 2003]{cant03} Cantiello, M., Raimondo, G., 
Brocato, E., \& Capaccioli, M.\ 2003, \aj, 125, 2783 

\bibitem[C\^ot\'e et al. 2004]{cot04}
C\^ot\'e, P., Blakeslee, J.P., Ferrarese, L., Jord\'an, A., Mei, S., Merritt, D., Milosavljevi\'c, M.,
Peng, E.W., \& West, M.J. 2004, \apjs, 153, 223 (Paper~I)

\bibitem[de Vaucouleurs 1948]{deVaucprof} de Vaucouleurs, G.\ 
1948, Ann.\ d'Astroph., 11, 247 
 
\bibitem[Ferrarese et al. 2000]{fer00a}
Ferrarese L., Mould, J.R., Kennicutt R.C. 2000, ApJ, 529, 745


\bibitem[Ford et al. 1998]{for98}
Ford, H.C. et al. 1998, Proc. SPIE, 3356, 234

\bibitem[Fruchter \& Hook 2002]{fru02} 
Fruchter, A.~S.~\& Hook, R.~N.\ 2002, \pasp, 114, 144 


\bibitem[Jensen et al. 1998]{jen98} Jensen, 
J.~B., Tonry, J.~L., \& Luppino, G.~A.\ 1998, \apj, 505, 111 

\bibitem[Jensen et al. 1999]{jen99}
Jensen, J.B., Tonry, J.L., Luppino, G.A., 1999, ApJ, 510, 71

\bibitem[Jensen et al. 2001]{jen01}
Jensen, J.B. et al., 2001, ApJ, 550, 503

\bibitem[Jensen et al. 2003]{jen03} Jensen, J.~B., Tonry, 
J.~L., Barris, B.~J., Thompson, R.~I., Liu, M.~C., Rieke, M.~J., Ajhar, 
E.~A., \& Blakeslee, J.~P.\ 2003, \apj, 583, 712 

\bibitem[Jerjen et al.2001]{jer01} 
Jerjen, H., Rekola, R., Takalo, L., Coleman, M., \& Valtonen, M.\ 2001, \aap, 380, 90 


\bibitem[Jerjen 2003]{jer03} 
Jerjen, H.\ 2003, \aap, 398, 63 

\bibitem[Jerjen et al.  2004]{jer04} 
Jerjen, H., Binggeli, B., \& Barazza, F.~D. 2004, \aj, 127, 771 

\bibitem[Jord\'an et al. 2004]{jor04}
Jord\'an, A., Blakeslee, J.P., Peng, E.W., Mei, S., C\^ot\'e, P., Ferrarese, L., Tonry, J.L., Merritt, D., Milosavljevi\'c, M., \& West, M.J. 2004, ApJS, in press (Paper~II)

\bibitem[Krist 2003]{kri03}
Krist, J. 2003, STScI Instrument Status Report ACS2003-06 

\bibitem[Lauer et al. 1998]{lau98}
Lauer T.R., Tonry, J.L., Postman, M. et al. 1998, ApJ, 499, 577          


\bibitem[Liu et al. 2000]{liu00}
Liu. M.C., Charlot, S., Graham G.R., 2000, ApJ, 543, 664

\bibitem[Liu \& Graham 2001]{liu01}
Liu, M.C. \& Graham J.R. 2001, ApJL, 557, 31

\bibitem[Liu et al. 2002]{liu02} 
Liu, M.~C., Graham, J.~R., \& Charlot, S.2002, \apj, 564, 216 

\bibitem[Luppino \& Tonry 1993]{lup93} 
Luppino, G.~A.~\& 
Tonry, J.~L.\ 1993, \apj, 410, 81 

\bibitem[Mei et al. 2000]{mei00} 
Mei, S., Silva, D., \& Quinn, P.~J. 2000, \aap, 361, 68 

\bibitem[Mei et al. 2001a]{mei01a} 
Mei, S., Silva, D.~R., \& Quinn, P.~J. 2001a, \aap, 366, 54 


\bibitem[Mei et al.  2001b]{mei01b}
Mei, S.,  Quinn, P.J., Silva, D.R. 2001b, A\&A,  371, 779

\bibitem[Mei et al. 2001c]{mei01c} 
Mei, S., Kissler-Patig, M., Silva, D.~R., \& Quinn, P.~J. 2001c, \aap, 376, 
793 

\bibitem[Mei et al.  2003]{mei03}
Mei, S.,  Scodeggio, M.,  Silva, D.R, Quinn, P.J. 2003, A\&A,  399, 441



\bibitem[Meurer et al. 2002]{meu02}     
Meurer, G., Lindler, D., Benitez, N., Blakeslee, J., Bouwens, R., Clampin, M., Cox, R.C. 2002,  Future EUV and UV Visible Space Astrophysics Missions and Instrumentation, eds. J. C. Blades \& O.H. Siegmund, Proc. SPIE, Vol. 4854, 507
   

\bibitem[Mieske \& Hilker 2003]{mie03a} 
Mieske, S.~\& Hilker, M.\ 2003, \aap, 410, 445 


\bibitem[Mieske et al. 2003]{mie03b}
 Mieske, S., Hilker, M., \& Infante, L.\ 2003, \aap, 403, 43 



\bibitem[Pahre \& Mould 1994]{pah94}
Pahre, M.A. \& Mould J.R. 1994, ApJ, 433, 567


\bibitem[Pahre et al. 1999]{pah99}
Pahre, M.A., Mould, J. R., Dressler, A. et al. 1999,  ApJ, 515, 79


\bibitem[Press et al. 1992]{press2}
 Press, W.H. et al. 1992,
{\it Numerical Recipes},  Cambridge University Press, New York


\bibitem[Sodemann \& Thomsen 1995]{sod95}
Sodemann, M. \& Thomsen, B., 1995, AJ, 110, 179

\bibitem[Sodemann \& Thomsen 1996]{sod96}
Sodemann, M. \& Thomsen, B., 1996, AJ, 111, 208


\bibitem[Thomsen et al. 1997]{tho97}
Thomsen, B., Baum, W.A., Hammergren M. {\it et al.}, 1997, ApJ, 483, L37



\bibitem[Tonry 1991]{ton91} Tonry, J.~L.\ 1991, \apjl, 373, L1 

\bibitem[Tonry \& Schneider 1988]{ts88}
Tonry, J.L. \& Schneider, D.P., 1988, AJ, 96, 807


\bibitem[Tonry et al. 1990]{ton90}
Tonry, J.L., Ajhar, E.A., Luppino G.A. 1990, AJ, 100, 1416


\bibitem[Tonry et al. 1997]{ton97}
Tonry, J.L., Blakeslee, J.P., Ajhar, E.A. {\it et al.} 1997, ApJ, 475, 399

\bibitem[Tonry et al. 2000]{ton00}
Tonry, J.L., Blakeslee, J.P., Ajhar, E.A., Dressler, A.
   2000, ApJ, 530, 625


\bibitem[Tonry et al. 2001]{ton01}
Tonry, J.L., Dressler,A.,   Blakeslee, J.P.  {\it et al.} 2001, ApJ, 546, 681 


\bibitem[worthey 1993]{wo93}
Worthey,G. 1993, ApJ, 409, 530 

\end{thebibliography}
\end{document}